\documentclass[aps,prd,nofootinbib,twocolumn,superscriptaddress,preprintnumbers]{revtex4-1}

\usepackage{amssymb}
\usepackage{amsmath}
\usepackage{epsfig}
\usepackage{hyperref}
\usepackage{breakurl}
\usepackage{bm}
\usepackage{color}
\usepackage{tikz}
\usepackage[utf8]{inputenc}
\usetikzlibrary{snakes}
\usetikzlibrary{decorations}
\usetikzlibrary{trees}
\usetikzlibrary{decorations.pathmorphing}
\usetikzlibrary{decorations.markings}
\usetikzlibrary{external}
\usetikzlibrary{intersections}
\usetikzlibrary{shapes,arrows}
\usetikzlibrary{arrows.meta}
\usetikzlibrary{calc}
\usetikzlibrary{shapes.misc}
\usetikzlibrary{decorations.text}
\usetikzlibrary{backgrounds}
\usetikzlibrary{fadings}
\usetikzlibrary{tikzmark,calc,arrows,shapes,decorations.pathreplacing}

\makeatletter
\def\simgt{\mathrel{\lower2.5pt\vbox{\lineskip=0pt\baselineskip=0pt
           \hbox{$>$}\hbox{$\sim$}}}}
\def\simlt{\mathrel{\lower2.5pt\vbox{\lineskip=0pt\baselineskip=0pt
           \hbox{$<$}\hbox{$\sim$}}}}

\def\fig#1{Fig.~\ref{#1}}

\makeatother

\def\eqn#1{Eq.~\eqref{#1}}
\def\spa#1.#2{\left\langle#1\,#2\right\rangle}
\def\spb#1.#2{\left[#1\,#2\right]}
\def\sand#1.#2.#3{%
\left\langle#1{\vphantom1}\right|{#2}\left|#3\right]}%
\def\sandmp#1.#2.#3{%
\left\langle#1{\vphantom1}\right|{#2}\left|#3\right]}%
\def\sandpm#1.#2.#3{%
\left[#1{\vphantom1}\right|{#2}\left|#3\right\rangle}%
\def\sandmm#1.#2.#3{%
\left\langle#1{\vphantom1}\right|{#2}\left|#3\right\rangle}%
\def\sandpp#1.#2.#3{%
\left[#1{\vphantom1}\right|{#2}\left|#3\right]}%

\def\Section#1{\noindent {\it #1}}
\newcommand{\be}{\begin{equation}}
\newcommand{\ee}{\end{equation}}

\renewcommand{\imath}{\mathrm{i}}

\newcommand{\sigmabar}{\bar \sigma}
\newcommand{\vect}{\boldsymbol}
\renewcommand{\imath}{\mathrm{i}}
\newcommand{\eikSum}{\mathcal{E}}

\def\topbotatom#1{\hbox{\hbox to 0pt{$#1\bot$\hss}$#1\top$}}

\allowdisplaybreaks

\begin{document}

\title{Second-order self-force potential-region binary dynamics at $\mathcal O(G^5)$ in supergravity}

\author{Zvi Bern}
\affiliation{
Mani L. Bhaumik Institute for Theoretical Physics,
University of California at Los Angeles,
Los Angeles, CA 90095, USA}

\author{Enrico Herrmann}
\affiliation{
	Mani L. Bhaumik Institute for Theoretical Physics,
	University of California at Los Angeles,
	Los Angeles, CA 90095, USA}

\author{Radu Roiban}
\affiliation{Institute for Theoretical Studies, ETH Zurich, 8092 Zurich, Switzerland}
\affiliation{Institute for Gravitation and the Cosmos and 
Institute for Computational and Data Sciences\\
Pennsylvania State University,
University Park, PA 16802, USA}

\author{Michael~S.~Ruf}
\affiliation{
Stanford Linear Accelerator Center, Menlo Park, United States
}

\author{Alexander V. Smirnov}
\affiliation{Research Computing Center, Moscow State University, 119991 Moscow, Russia}
\affiliation{Moscow Center for Fundamental and Applied Mathematics, 119992 Moscow, Russia}

\author{Vladimir A. Smirnov}
\affiliation{Moscow Center for Fundamental and Applied Mathematics, 119992 Moscow, Russia}
\affiliation{Skobeltsyn Institute of Nuclear Physics of Moscow State University, 119991, Moscow, Russia}

\author{Mao Zeng}
\affiliation{Higgs Centre for Theoretical Physics, University of Edinburgh, Edinburgh, EH9 3FD, United Kingdom}

\begin{abstract}

We compute the potential-graviton contributions to the conservative scattering angle of two non-spinning bodies in maximal supergravity at fifth order in Newton's constant, including second-order self-force effects. Our goal is to tackle the challenging integrals arising at this order in Einstein gravity, but within the technically simpler framework of supergravity. The calculation employs the scattering-amplitude framework, effective field theory, and multi-loop integration techniques based on integration by parts and differential equations. The final result is expressed as a series expansion around the static limit, thereby avoiding the explicit evaluation of intricate special functions. This series solution for the master integrals applies, as well, to the corresponding computation in general relativity. Remarkably, we observe nontrivial cancellations among contributions associated with Calabi--Yau integrals, alongside a distinct contribution governed by a Heun differential equation.
\end{abstract}
   
\maketitle

\Section{Introduction---}
\label{sec:intro}
%
The detection of gravitational waves~\cite{LIGOScientific:2016aoc, LIGOScientific:2017vwq} marked a milestone in physics, opening a new window into the universe.  Upcoming improvements in gravitational-wave detectors~\cite{Punturo:2010zz, LISA:2017pwj, Reitze:2019iox, LIGOasharp, Abac:2025saz} anticipate dramatic gains in both sensitivity and frequency coverage. Realizing the full scientific potential of these observations, however, demands high-precision theoretical modeling across a large parameter space. Meeting this formidable challenge will require significant advances in multiple complementary frameworks, including numerical relativity~\cite{Pretorius:2005gq, Campanelli:2005dd, Baker:2005vv, Damour:2014afa},  the gravitational self-force (SF) program~\cite{Mino:1996nk, Quinn:1996am, Poisson:2011nh, Barack:2018yvs}, effective field theory (EFT)~\cite{Goldberger:2004jt, Cheung:2018wkq}, and both post-Newtonian (PN)~\cite{Droste:1916, Droste:1917, Einstein:1938yz, Ohta:1973je, Blanchet:2013haa} and post-Minkowskian (PM)~\cite{Bertotti:1956pxu, Kerr:1959zlt, Bertotti:1960wuq, Westpfahl:1979gu, Portilla:1980uz, Bel:1981be} methods. In the end, advances in the various approaches must be synthesized into accurate and robust waveform models, such as those within the effective-one-body (EOB) framework~\cite{Buonanno:1998gg, Buonanno:2000ef}.  

The PM expansion in powers of Newton's constant $G$ is the natural perturbative framework for scattering problems with large minimal separation, as well as for bound motion on eccentric orbits~\cite{Khalil:2022ylj}. By preserving manifest Lorentz symmetry, relativistic quantum field theory and scattering-amplitude methods provide a systematic, gauge-invariant route to higher-order PM dynamics, which can also include finite-size, tidal, spin, and environmental effects (see e.g. Refs.~\cite{Blanchet:2013haa, Porto:2016pyg, Buonanno:2022pgc, Barausse:2014tra}). This approach builds on important advances in the understanding of scattering amplitudes and in the ability to compute them. They include generalized unitarity~\cite{Bern:1994zx, Bern:1994cg, Bern:1997sc, Britto:2004nc, Bern:2004cz, Bern:2007ct}, the double-copy construction~\cite{Kawai:1985xq, Bern:2008qj, Bern:2010ue, Bern:2019prr}, and powerful integration methods~\cite{Chetyrkin:1981qh, Tkachov:1981wb, Kotikov:1990kg, Bern:1993kr, Remiddi:1997ny, Gehrmann:1999as, Henn:2013pwa}.
A number of field-theory formalisms have been developed, including those based on matching to effective field theory~\cite{Cheung:2018wkq}, eikonal exponentiation~\cite{DiVecchia:2020ymx, DiVecchia:2021bdo}, an exponential representation of the S-matrix~\cite{Damgaard:2023ttc}, an observable-focused setup~\cite{Kosower:2018adc}, heavy-particle effective theories~\cite{Damgaard:2019lfh}, extreme mass-ratio EFT~\cite{Cheung:2023lnj, Kosmopoulos:2023bwc}, and classical worldline methods~\cite{Kalin:2020fhe, Mogull:2020sak, Kalin:2022hph, Jakobsen:2022psy}---and were applied to the gravitational two-body problem at higher PM orders, see e.g., Refs.~\cite{Bern:2019crd, Bern:2019nnu, Bern:2024adl, Bern:2021dqo, Bern:2021yeh, Driesse:2024xad, Driesse:2024feo, Jakobsen:2023hig, Jakobsen:2023ndj, Dlapa:2021vgp, Dlapa:2021npj, Bjerrum-Bohr:2021wwt} for spinless configurations, with the current frontier being to complete the fifth order, ${\cal O}(G^5)$, 5PM.

In this paper, we address a key bottleneck towards achieving 5PM-order accuracy for the two-body problem in Einstein's gravity, organized following the gravitational self-force mass-ratio expansion~\cite{Damour:2019lcq}.  Substantial recent progress~\cite{Bern:2023ccb, Bern:2024adl, Driesse:2024feo, Frellesvig:2024rea, Brammer:2025rqo, Duhr:2025lbz} includes the computation of the conservative 5PM sector through 1SF~\cite{Driesse:2024xad}. Although the 1SF correction to the probe limit suffices to determine the scattering angle through 4PM, at 5PM it receives a 2SF contribution for the first time.
The central difficulty lies in evaluating the loop integrals entering the 2SF sector. Integration-by-parts (IBP) reduction~\cite{Chetyrkin:1981qh, Tkachov:1981wb} of high tensor-rank integrals to a relatively small number of master integrals poses a considerable challenge. Here we use a modified version of FIRE7~\cite{FIRE7} for this step.  Subsequently, the master integrals are evaluated by differential-equation methods~\cite{Kotikov:1990kg, Bern:1993kr, Remiddi:1997ny, Gehrmann:1999as, Henn:2013pwa}. At present, these are the only viable methods capable of handling the relevant integrals. However, they require substantial computer resources as well as advanced algorithms, such as finite-field methods~\cite{vonManteuffel:2014ixa, Peraro:2016wsq}.

Because the 2SF integrals are highly nontrivial, we instead work in the simpler setting of $\mathcal{N}=8$ supergravity. This theory retains the essential features of general relativity, including the same integral families and master integrals, and allows us to carry over solutions directly to the gravitational case. The key advantage is that the most complicated integrals have a lower tensor rank than in Einstein's gravity, enabling quicker access to gravitational 2SF results.  Supersymmetric setups have long served as effective laboratories for gravitational calculations~\cite{Caron-Huot:2018ape, Parra-Martinez:2020dzs, DiVecchia:2020ymx, DiVecchia:2021bdo} and, in particular, have already been used to study the 1SF integrals in Ref.~\cite{Bern:2024adl}.

\medskip
\Section{From Quantum Amplitudes to Classical Observables---}
%
At large distances, black holes can be modeled as massive point particles, and their post-Minkowskian dynamics can be captured by the classical limit of quantum scattering amplitudes~\cite{Cheung:2018wkq, Kosower:2018adc, DiVecchia:2020ymx, DiVecchia:2021bdo, Bern:2021dqo, Damgaard:2023ttc}. By the correspondence principle, this limit amounts to an expansion at large angular momentum, $J \gg \hbar$, or equivalently at small momentum transfer, $q = p_1 + p_4$.\footnote{All momenta are taken incoming: $p_1{+}p_2{+}p_3{+}p_4{=}0$.} Each term depends nontrivially on the dimensionless boost parameter $\sigma = p_1 \cdot p_2 /(m_1 m_2)=\frac{1}{\sqrt{1-v^2}}$, or equivalently the relative velocity $v$. The classical limit further selects loop momenta of order $|q|\ll m_i, \sqrt{s}$ so that, within the EFT-based \emph{method of regions}~\cite{Beneke:1997zp}, the relevant contributions come from the soft region. At ${\cal O}(G^5)$ this procedure yields a classical amplitude ${\cal M}^{\text{cl}}_5$ proportional to $q^2 \ln(-q^2)$, corresponding to a $1/r^5$ potential.

We focus on the \emph{potential region}, the subset of the soft region describing instantaneous interactions between massive particles, where loop momenta scale as $\ell = (\ell^0,\boldsymbol{\ell}) \sim |q|(v,\mathbf{1})$ with $v \ll 1$. This region suffices to capture the challenging integrals. Classical observables are then extracted from the \emph{radial action}, which is related to the massive-scalar $2 \to 2$ scattering amplitude~\cite{Bern:2021dqo} by,
\begin{equation}
\label{aarelation}
i \mathcal M^\text{cl}(\bm q) = \int_J \big(e^{i I_r(J)} - 1\big)\,, 
\end{equation}
where the integration measure will be defined in \eqn{eq:FT_J} below. The classical radial action~\cite{Landau:1975pou}, $I_r(J)=\int_\gamma p_r\, \mathrm{d}r$, is an integral over the radial momentum $p_r$ along the trajectory $\gamma$. It corresponds to the classical action with fixed total energy $E$ and angular momentum $J$ and determines the scattering angle
\begin{equation}
\label{eq:angle_Irad}
\chi= - \frac{\partial I_r(J)}{\partial J} 
         = \pi - 2 J \int\limits_{r_{\mathrm{min}}}^\infty \frac{\mathrm{d}r}{r^2 \sqrt{p^2_r(r)}}\,.
\end{equation}
In the PM expansion, the amplitude, the radial action, and the scattering angle are series in Newton's constant,
\begin{equation}
\label{eq:angle_PM}
    \mathcal{M}=\sum_{k=1}\mathcal{M}_k\,,\hskip.4cm \tilde{I}_r=\sum_{k=1}\tilde{I}_{r,k}\,,\hskip.4cm 
    \chi=\sum_{k=1}\chi_k\, ,
 \end{equation}
where $\mathcal{M}_k=\mathcal{O}(G^k)\,,\ \tilde{I}_{r,k}=\mathcal{O}(G^k)\,,$ and $\chi_k=\mathcal{O}(G^k)$ truncate exactly at the classical order, and we leave implicit their kinematic dependence. At the fifth order
 \begin{align}
 {\cal M}_5(\bm q) ={}& 
            \tilde{I}_{r,5}(\bm q) 
          + \mathcal{M}_5^{{\rm it.}}(\bm q) \ ;
 \label{eq:amp_action_5}
\end{align} 
$\mathcal{M}_5^{{\rm it.}}$ contains all the classically-singular terms, see e.g. Ref.~\cite{Bern:2024adl} for details. As noted there, such terms are separated by the boundary conditions in the differential equations for the master integrals. The classical part of the $2{\to}2$ amplitude, i.e.~the radial action, is
\begin{equation}
\tilde I_r (\bm q) \, {=} 
\int_{J}\! I_r(J) :=
4E\, |\vect{p}|\,  \mu^{-2\epsilon}\!\!\int {\mathrm{d}^{D-2}\bm b}\, e^{\imath\bm q\cdot \bm b}\, I_r(J) \, .
\label{eq:FT_J}
\end{equation}
In the center-of-mass frame, the asymptotic impact parameter $\vect{b}$ and the momentum transfer~$\vect{q}$ are Fourier conjugate; $\vect{p}$ is the spatial momentum.
We use dimensional regularization, with $D{=}4{-}2\epsilon$ and regularization scale $\mu$. The energy $E$, $|\vect{p}|$ and $\sigma$ are related as $E^2 = m^2_1 {+} m^2_2 {+} 2 m_1m_2\, \sigma\,, \text{and } |\vect{p}| {=} \frac{m_1 m_2}{E} \sqrt{\sigma^2 {-}1}\,.$ The 5PM radial action is organized by independent mass structures
\begin{align}
\label{eq:5PM_amp_SF_org}
    \frac{\tilde{I}_{r,5}}{m^4_1 m^4_2} ={}& 
        \frac{(m_1^4{+}m_2^4)}{m_1^2 m_2^2} \tilde{I}_{r,5}^{0 {\rm SF}} 
    + \frac{(m_1^2{+}m_2^2)}{m_1 m_2}    
        \tilde{I}_{r,5}^{1 {\rm SF}}
    + \tilde{I}_{r,5}^{2 {\rm SF}} 
   \\[3pt]  
    = {}&
       \frac{M^8\, \nu^2}{m_1^4 m_2^4}
            \left[ 
                      \hat{I}_{r,5}^{0 {\rm SF}} 
                +\nu  \hat{I}_{r,5}^{1 {\rm SF}}  +\nu^2\hat{I}_{r,5}^{2 {\rm SF}}
            \right]
   , \nonumber
\end{align}
where we define $M{=}m_1{+}m_2$, $\nu {=} \frac{m_1m_2}{M^2}$,  $\hat{I}_{r,5}^{0 {\rm SF}} {=} \tilde{I}_{r,5}^{0 {\rm SF}} $, 
$\hat{I}_{r,5}^{1 {\rm SF}} {=} \tilde{I}_{r,5}^{1 {\rm SF}}{-}4\tilde{I}_{r,5}^{0 {\rm SF}}$ and 
$\hat{I}_{r,5}^{2 {\rm SF}} {=} \tilde{I}_{r,5}^{2 {\rm SF}}{-}2\tilde{I}_{r,5}^{1 {\rm SF}}{+}2\tilde{I}_{r,5}^{0 {\rm SF}}$.
While the first form in \eqn{eq:5PM_amp_SF_org} is more natural from an amplitudes perspective, the second $\nu$-dependent form is more standard for the self-force expansion.

\medskip
\Section{Computational Setup---}
%
We extract classical conservative 5PM observables from the classical scattering amplitude of massive particles at $\mathcal{O}(G^5)$, i.e.~at four-loop order.

On-shell techniques have proven useful for deriving loop integrands for classical massive scattering amplitudes, including in general relativity.
Here we follow Refs.~\cite{Caron-Huot:2018ape, Parra-Martinez:2020dzs, Bern:2024adl} and construct the massive-scalar $2\,{\to}\,2$ amplitude integrand at four loops in $\mathcal{N}=8$ supergravity by dimensionally reducing the corresponding higher-dimensional massless supergravity integrand.
These higher-dimensional amplitudes were constructed in Ref.~\cite{Bern:2012uf, Bern:2018jmv}, through ${\cal O}(G^6)$ using the generalized unitarity method~\cite{Bern:1994zx, Bern:1994cg, Bern:1997sc, Britto:2004nc, Bern:2004cz, Bern:2007ct} in conjunction with the double copy~\cite{Kawai:1985xq, Bern:2008qj, Bern:2010ue, Bern:2019prr}. 
Upon an appropriate dimensional reduction from six or higher dimensions, the components of the external momenta in the extra dimensions act as independent masses for the four-dimensional scalar fields. 
Furthermore, we retain the dependence on the BPS-angle through the variable $c_\phi \equiv \cos\phi-\sigma$. 

\begin{figure}[tb]
$
\vcenter{\hbox{
\includegraphics[scale=0.9]{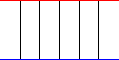}}} 
 \hskip .2 cm 
\vcenter{\hbox{\includegraphics[scale=0.9]{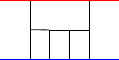}}} \hskip .2 cm 
\vcenter{\hbox{\includegraphics[scale=0.9]{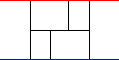}}} \hskip .2 cm 
\vcenter{\hbox{\includegraphics[scale=0.9]{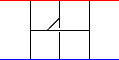}}}
$
\caption{Four-loop sample diagrams contributing through 2SF.}
\label{fig:egFeynmanDiags}
\end{figure}

As discussed above, we extract the classical potential-region contributions to the massive scalar amplitude by expanding it up to ${\cal O}(|\vect{q}|^2)$. 
Upon expansion, the complete $|\vect{q}|^2$ and $m_i$ dependence can be separated so that all integrals are functions of the single kinematic variable $\sigma$.
The even and odd powers of $|\vect{q}|$ are decoupled, so, without affecting the classical contributions, we will restrict to terms that contain only even powers of $|\vect{q}|$.
The result is organized diagrammatically in terms of 394 integral families with only cubic vertices. Sample diagrams are shown in Fig.~\ref{fig:egFeynmanDiags}. Notably, the third and fourth diagrams appear for the first time at 2SF order, whereas the first two were present already in previous 1SF computations. All matter lines represent linearized (eikonal) propagators. Compared to classical worldline methods, our setup involves more integral families. For all but the most complicated 2SF cases, the extra integrals are, however, computationally subleading. 
To eliminate remaining undesirable redundancies in the integrand, we map it to a global basis following Ref.~\cite{Bern:2024vqs}. 

Given an integrand, the main bottleneck is loop integration. As usual, we first IBP-reduce the $\mathcal{O}(10^6)$ integrals of the classical amplitude to a set of $\mathcal{O}(10^3)$ \emph{master integrals}. These are chosen to prevent non-factorized dependence on kinematic variables and the spacetime dimension in denominators~\cite{Smirnov:2020quc, Usovitsch:2020jrk}; integrals with up to 4 units of complexity (defined as the sum of the tensor rank and of the propagators' powers above 1) are needed in some 2SF integral families to realize this property. 
We use FIRE7~\cite{FIRE7} with enhanced performance for finite fields~\cite{vonManteuffel:2014ixa, Peraro:2016wsq} and automatic MPI-parallelized reduction and reconstruction of analytic master integral coefficients.
The most difficult integral families are reduced with an experimental FIRE setup with improved seeding, which was implemented outside FIRE in Ref.~\cite{Bern:2024adl} (see also Refs.~\cite{Driesse:2024xad, Guan:2024byi, Lange:2025fba}). 
In $\mathcal N=8$ supergravity, supersymmetry lowers the tensor rank of the most difficult integrals by four units relative to Einstein's gravity, greatly simplifying IBP reduction and providing a faster path to examining 2SF results in a gravitational theory. 

The difficulty of the IBP reduction depends on several key factors; chief among them are the complexity of the integrals,  the topology of the diagram, and the number of master integrals at the top level.  The non-planar classical integral topologies, especially for families that only contribute to classical parts at 2SF, are particularly challenging and a dominant part of the calculation. 
Because of this, we \emph{planarize} the most complicated integrand contributions, which arise from permuting vertices on the matter lines in the third and fourth diagrams in Fig.~\ref{fig:egFeynmanDiags}, by judiciously using the support of cut matter lines~\cite{Saotome:2012vy, Akhoury:2013yua} following the discussions in Sec.~3.2 of Ref.~\cite{Bern:2024adl}.  

Having completed the IBP reduction of the classical four-loop amplitude, the result is expressed in terms of approximately $4{\small,}000$ master integrals that are functions of the spacetime dimension $D=4{-}2\epsilon$, and of $\sigma$. They are evaluated with the method of differential equations~\cite{Kotikov:1990kg, Bern:1993kr, Remiddi:1997ny, Gehrmann:1999as, Henn:2013pwa}. In the most condensed form, all master integrals are collected into a single vector $\vec{I}$ and differential equations are derived with the help of IBP identities,
\begin{align}
\label{eq:diff_eq}
    \frac{\mathrm{d}}{\mathrm{d}\sigma} \vec{I}(\sigma,\epsilon) = \hat{M}(\sigma,\epsilon) \vec{I}(\sigma,\epsilon)\, ,
\end{align}
subject to appropriate boundary conditions. In our case, it is convenient to supply boundary conditions in the static $\sigma=1$ limit. 
The form of the connection matrix $\hat{M}(\sigma,\epsilon)$ depends on the initial choice of basis $\vec{I}(\sigma,\epsilon)$, but is guaranteed to be in a block-triangular form, with diagonal blocks representing the differential equations of the relevant sectors, and off-diagonal couplings between parent and daughter sectors.\footnote{A \emph{sector} is a subset of Feynman integrals where the same propagators appear with a positive power in the denominator.}
In cases where the solutions to Eq.~\eqref{eq:diff_eq} can be written in terms of (generalized) polylogarithms~\cite{Goncharov:1998kja}, it is convenient to rotate the basis, $\vec{I}$, to bring the differential equation into \emph{canonical} form~\cite{Henn:2013pwa}, where the dependence on the spacetime dimension via $\epsilon$ factors out, and the matrix simplifies $\hat{M}_{{\rm can.}}(\sigma,\epsilon) = \epsilon \sum_\alpha \hat{M}_\alpha \, \tfrac{\mathrm{d}}{\mathrm{d}\sigma}\log w_\alpha(\sigma)$. The $w_\alpha(\sigma)$ are the \emph{letters} and the entries of the matrices $\hat{M}_\alpha$ are rational numbers. In this form, the solutions of the differential equations are given in terms of Chen iterated integrals over logarithmic integration kernels to the desired order in the $\epsilon$ expansion. These integrations are algorithmic and are implemented in various codes, e.g.~\cite{Panzer:2014caa, Duhr:2019tlz}. 

There are significant ongoing efforts to extend such methods to elliptic integrals~\cite{Adams:2018yfj, Gorges:2023zgv}, and Calabi--Yau geometries~\cite{Frellesvig:2023bbf, Klemm:2024wtd, Frellesvig:2024rea, Duhr:2025lbz}, which not only appear in the classical binary problem, but also in amplitudes with internal masses relevant for particle-physics applications, see e.g.~\cite{Bourjaily:2022bwx}. Despite the recent progress, solving differential equation systems for the complicated underlying geometries still requires a case-by-case analysis. 


We take a different approach here. Instead of seeking a closed-form solution in terms of special functions, we first note that in the potential region the master integrals are meromorphic functions in $\sigmabar=\sigma{-}1$ near $\sigmabar=0$ and therefore locally admit a Laurent-series representation about the static limit.\footnote{In the potential region, the meromorphicity follows from potential-region scaling of the loop momenta $\ell^\mu = (\omega, \bm \ell) \sim (v \lambda, \lambda)$ which leads to an expansion containing only integer powers of $v$. In other regions, e.g.~the radiation region, the scaling is different and factors of the form $\sigmabar^{-\alpha \epsilon}$ appear. See, e.g., Refs.~\cite{Bern:2019crd,  Bern:2021dqo,  Bern:2021yeh}.} We thus solve the differential equations by the Frobenius method (see e.g.~Refs.~\cite{Mistlberger:2018etf, Moriello:2019yhu, Pozzorini:2005ff, Hidding:2020ytt} for applications in collider physics), of the form
\begin{align}
\label{eq:diff_eq_sol}
    \vec{I}(\sigma,\epsilon) = \sum^{n_{{\rm max}}}_{n=n_0} \sigmabar^n\, \vec{c}_n(\epsilon) + \mathcal{O}(\sigmabar^{n_{{\rm max}}{+}1}) \,.
\end{align}
The coefficients $\vec{c}_n(\epsilon) $ are rational functions of the dimensional regulator $\epsilon$.
The lower limit $n_0$ can be inferred from the leading velocity scaling of the master integrals in the static limit. The upper limit $n_{{\rm max}}$ is fixed by the highest desired order of the solution. 

Combining Eq.~\eqref{eq:diff_eq_sol} with a $\sigmabar$ expansion of $\hat{M}(\sigma,\epsilon)$, which has poles up to eighth order at $\sigmabar=0$, we solve the differential system order by order in $\sigmabar$.
At each order, the coefficients $c_{n,j}$ are determined by a linear system; we 
rationally-reconstruct~\cite{Wang:1981, Wang:1982} the $\epsilon$-dependence 
of their solutions from multiple numerical points over prime fields.
We construct the solution through $\mathcal{O}(\sigmabar^{15})$, but going to higher orders is, of course, possible. 

Having planarized the most complicated classical 2SF diagram topologies, we find 371 boundary conditions that are identical in maximal supergravity and general relativity. They are linked to the static limit of scalar integrals encompassing both classically singular and regular contributions; in this limit, their evaluation is relatively simple. Notably, we recover all the 312 that appeared at 1SF order~\cite{Bern:2024adl}, thus having a strong consistency check.

\medskip
\Section{Results---}
%
Individually, these abstract static boundary integrals are ill-defined without additional regularization. However, upon inserting the series solution for the master integrals $\vec{I}(\sigma,\epsilon)$ through $\mathcal{O}(\sigmabar^{15})$ into the IBP-reduced classical $\mathcal{N} = 8$ supergravity amplitude, they organize into well-defined combinations---\emph{eikonal sums}~\cite{Saotome:2012vy, Akhoury:2013yua}---which we denote by ${\cal E}$. All diagrams within a given sum are related by permuting vertices on matter lines, forcing them to be cut. The ability to planarize the classical 2SF diagram topologies, as discussed above, relies on the emergence of such combinations. In terms of eikonal sums, the $|\bm q|$-expanded potential-region amplitude through classical order has the form
\begin{equation}
\label{eq:amp_sterman_organized}
    \mathcal{M}_5=\left(\frac{|{\vect{q}}|^2}{\mu^2}\right)^{-4\epsilon}
    \sum_{k=-2}^2|{\vect{q}}|^{k}
    \sum_{n=1}^{N_k}
    f_{k,n}(\sigmabar,\epsilon) \, \eikSum_{k,n}(\epsilon)
    \,,
\end{equation}
where $N_k$ counts the independent eikonal sums and depends on the power, $k$, of $|{\vect{q}}|$. As mentioned above, we focus on the even-in-$|{\vect{q}}|$ terms. The functions $f_{k,n}(\sigmabar,\epsilon)$ are given as series in the velocity variable $\sigmabar$, and we suppress their mass and BPS-angle dependence. Retaining the computationally simpler classical iterations serves as a non-trivial check of our setup. We have verified that the $k=-2,0$ terms agree with our previous results in~\cite{Bern:2024adl} and refrain from repeating them.
We focus instead on the classical contributions $\tilde{I}_{r,5}$, obtained from $\mathcal{M}_5$ by dropping iteration terms, c.f.~Eq.~\eqref{eq:amp_action_5}. This corresponds to taking the $k=2$ term in Eq.~\eqref{eq:amp_sterman_organized}. In order to generate a classical contribution, the combination $f_{2,n}(\sigmabar,\epsilon) \, \eikSum_{2,n}(\epsilon)$ needs to develop poles in $\epsilon$. At the classical order, there are 16 eikonal sums---one in the 0SF sector, six in the 1SF sector, and nine at 2SF. The 0SF and 1SF sums have already been discussed in Sec.~5.2 of Ref.~\cite{Bern:2024adl}, so we give representative examples
\begin{align}
\eikSum_{2,12} = \vcenter{\hbox{\includegraphics[scale=0.7]{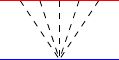}}}
+ \cdots \,,
\quad 
\eikSum_{2,1} {=} 
\hspace{-.1cm}
\vcenter{\hbox{\includegraphics[scale=0.7]{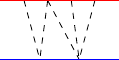}}}
\hspace{-.1cm}
+ \cdots \,.
\end{align}
The dashed lines denote spatial propagators, and the sums reduce to three-dimensional Euclidean integrals upon evaluating the energy integral in each loop. The remaining nine 2SF eikonal sums are, 
\begin{align}
\eikSum_{2,2} & = \vcenter{\hbox{\includegraphics[scale=0.7]{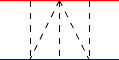}}}
+ \cdots \,,
\quad 
\eikSum_{2,3} = \vcenter{\hbox{\includegraphics[scale=0.7]{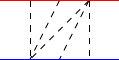}}}
+ \cdots \,,
\nonumber \\
\eikSum_{2,5} & = \vcenter{\hbox{\includegraphics[scale=0.7]{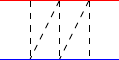}}}
+ \cdots \,,
\quad 
\eikSum_{2,7}  = \vcenter{\hbox{\includegraphics[scale=0.7]{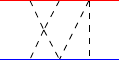}}}
+ \cdots \,,
\nonumber \\
\eikSum_{2,8} & = \vcenter{\hbox{\includegraphics[scale=0.7]{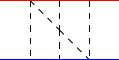}}}
+ \cdots \,,
\quad
\eikSum_{2,10}  = \vcenter{\hbox{\includegraphics[scale=0.7]{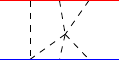}}}
+ \cdots \,,
\nonumber \\
\eikSum_{2,14} & = \vcenter{\hbox{\includegraphics[scale=0.7]{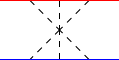}}}
+ \cdots \,,
\quad
\eikSum_{2,15}  = \vcenter{\hbox{\includegraphics[scale=0.7]{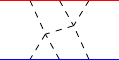}}}
+ \cdots \,,
\nonumber \\
\eikSum_{2,16} & = \vcenter{\hbox{\includegraphics[scale=0.7]{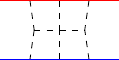}}}
+ \cdots \,.
\end{align}
Three of them, $\mathcal{E}_{2,2}$, $\mathcal{E}_{2,3}$, and $\mathcal{E}_{2,5}$, have already appeared in the 2SF  result for classical electrodynamics~\cite{Bern:2023ccb}. Having exploited the emergence of eikonal sums in the planarization of classical 2SF integrals, we treat all analogous terms by identifying the corresponding scalar boundary integrals with their counterpart in which all matter lines are cut.
In all terms independent of the planarization procedure (i.e., those proportional to $c^{n\geq 3}_\phi$), the eikonal sums emerge as expected and are a check of our integration procedure. 

Interestingly, in $\mathcal{N}=8$ supergravity, we find that the coefficients of three classical 2SF eikonal sums---$\mathcal{E}_{2,7}$, $\mathcal{E}_{2,8}$, and $\mathcal{E}_{2,15}$---vanish up to a sufficiently high order in $\epsilon$ so they do not contribute to the scattering angle.
Two of them tag interesting classes of non-polylogarithmic integrals.
The $\mathcal{E}_{2,8}$ sector is associated with a Heun-type~\cite{MRAmplitudes2023}, or special symmetric-square K3~\cite{Brammer:2025rqo} Picard--Fuchs equation, while $\mathcal{E}_{2,15}$ corresponds to a Calabi--Yau threefold~\cite{Brammer:2025rqo, Frellesvig:2024rea, Duhr:2025lbz}. 
Their vanishing contribution, reminiscent of a similar cancellation in the 4PN $\mathcal{O}(G^5)$ term of Einstein's gravity~\cite{Foffa:2016rgu}, makes it natural to conjecture that these integrals will be absent from the final result.
Similar cancellations are observed for the elliptic functions at 4PM in $\mathcal{N}=8$ and gravitational theories at 5PM 1SF~\cite{Bern:2024adl, Driesse:2024xad}. Lastly, although $\mathcal{E}_{2,7}$ is not tied to a special geometry, it likewise does not contribute to classical observables in $\mathcal{N}=8$ supergravity. 
After evaluating the eikonal sums and adding all contributions, we find 
\begin{align}
 \label{eq:resultnSF}
 \hspace{-.4cm}
    \tilde{I}_{r,5}^{n {\rm SF}}{=}
    G^5 \pi  c_{\phi }^4 \, 
     |{\vect{q}}|^2 \! 
     \left[
        \frac{|{\vect{q}}|^2}{\bar{\mu}^2}
     \right]^{\!-4\epsilon}\!\!
    \left[
        \frac{\tilde{I}_{r,5}^{{\rm nSF,div.\!\!\!\!}}}{\epsilon^2} 
        {+} \frac{\tilde{I}_{r,5}^{{\rm nSF,fin.\!\!\!\!}}}{\epsilon} 
        {+} \mathcal{O}(\epsilon^0)
    \right]\!, \,
  \hspace{-.4cm}  
\end{align}
where $ \bar \mu^ 2 =  4\pi e^{-\gamma_{\rm E}}\,\mu^2$ is the $\overline{\text{MS}}$ renormalization scale. The non-analytic $|{\vect{q}}|^2\ln|{\vect{q}}|$ term encoding the classical physics at 5PM, is obtained by expanding Eq.~(\ref{eq:resultnSF}) in $\epsilon$, i.e. $\tilde{I}_{r,5}^{{\rm nSF,fin.}}$ is the term accompanying the single-pole in $\epsilon$. 
The $1/\epsilon^2$ pole is due to the overlap between the potential and radiation contributions~\cite {Manohar:2006nz, Porto:2017dgs}. The divergence should cancel once tail contributions are included, leaving behind a term with logarithmic dependence on $\sigmabar$.  This cancellation first occurs at three loops~\cite{Bern:2021yeh} and is, in fact, a generic feature of the method of regions~\cite{Beneke:1997zp}.

Although we solved the differential equations through $\mathcal{O}(\sigmabar^{15})$, due to various poles in $\sigmabar$ in the amplitude coefficients, we obtain only the first eleven orders of the classical radial action around the static limit $\sigmabar\to 0$, which corresponds to the 10PN terms at $G^5$.
For ease of comparison with Ref.~\cite{Bern:2024adl}, we keep the dependence on $c_\phi$. 
At 0SF, we find $\tilde{I}_{r,5}^{0 {\rm SF,div.}} = 0$, and 
\begin{align}
\hspace{-.4cm}
    \tilde{I}_{r,5}^{0 {\rm SF,fin.}} {=} 
    \left(\!
    {-}\frac{1}{20 \sigmabar^4}
    {+}\frac{1}{10 \sigmabar^3}
    {-}\frac{1}{8 \sigmabar^2}
    {+}\cdots\!
   \right) c_{\phi }^6\,,
   \label{eq:series_0SF}
\hspace{-.4cm}   
\end{align}
where higher-order terms in the $\sigmabar$ expansion are suppressed. We provide them in the supplementary file \texttt{N8\textunderscore 2SF\textunderscore classical\textunderscore momentum\textunderscore space\textunderscore radial\textunderscore action.m} through $\mathcal{O}(\sigmabar^6)$ and checked that Eq.~\eqref{eq:series_0SF} agrees with the expansion of Eq.~(5.30) in Ref.~\cite{Bern:2024adl}, $\tilde{I}_{r,5}^{0 {\rm SF,fin.}} {=}-{4 c_{\phi }^6}/{5 \left((\sigmabar+1)^2-1\right)^4}$ on the overlap. 

The divergent part of the 1SF radial action is
\begin{align}
\tilde{I}_{r,5}^{1 {\rm SF,div.}}  = 16 c^2_\phi \Bigg[
& c^2_\phi \left(
            {-}\frac{1}{6 \sigmabar}
            {+}\frac{1}{20}
            {+}\frac{19 \sigmabar}{840}
            {+}\cdots
         \right)
\\ & \hspace{-2.2cm}        
{-} c_\phi \left(
            \frac{1}{\sigmabar}
            {+}\frac{5}{6}
            {-}\frac{49 \sigmabar}{60}
            {+}\cdots\!
        \right) 
{-} \left(
        4 
        {-}\frac{4 \sigmabar}{3} 
        {+}\frac{8 \sigmabar^2}{15}
        {+} \cdots \! 
        \right) \!
\Bigg]\,, \nonumber 
\end{align}
which agrees with the expansion of Eq.~(5.32) in Ref.~\cite{Bern:2024adl} to the order computed here. The finite part of the radial action, $\tilde{I}_{r,5}^{1 {\rm SF,fin.}}$, contains six $c_\phi$ structures and is too lengthy to print here, but is provided in the supplementary file. It also agrees with our earlier exact-in-velocity result for the overlapping orders.

The key new result of this work is the 2SF contribution to the classical radial action, which determines the corresponding potential-region scattering angle. 
Interestingly, the $1/\epsilon^3$ singularity of the three eikonal sums $\mathcal{E}_{2,2}$, $\mathcal{E}_{2,3}$, and $\mathcal{E}_{2,5}$ turns out to be spurious. It cancels in the 
full sum, leaving a divergent $1/\epsilon^2$ term,
\begin{align}
\tilde{I}_{r,5}^{2 {\rm SF,div.}} = 16 c^2_\phi \Bigg[
& 
    c^2_\phi  \left( 
        {-}\frac{1}{3 \sigmabar}
        {+}\frac{19}{90}
        {-}\frac{41 \sigmabar}{1260}
        {+} \cdots
    \right)
\\ 
& \hspace{-2.7cm}
{-}c_\phi \left(       
        \frac{2}{\sigmabar}
        {+}1
        {-}\frac{173 \sigmabar}{90}
        {+}\cdots\!
    \right)
{-} \left(  
        8 
        {-}\frac{16 \sigmabar}{3}
        {+}\frac{136 \sigmabar^2}{45} 
        {+} \cdots\!
    \right)
\Bigg]\,. \nonumber 
\end{align}
The presence of this 2SF divergence highlights new features, extending beyond leading order, in the relation between the lower-loop energy loss and the tail divergence.
Comparing the series solutions of the divergent parts of the 1SF and 2SF radial actions through ${\cal O}(\bar\sigma{}^5)$, we find the surprising relation,
\begin{align}
\label{eq:strange_relation}
\tilde{I}_{r,5}^{2 {\rm SF,div.}} = 
\frac{4}{\sqrt{\sigmabar}}\,{\rm arcsinh}\left(\sqrt{\sigmabar/2}\right) \, \tilde{I}_{r,5}^{1 {\rm SF,div.}} \,,
\end{align}
which is confirmed through the sixth order, indicating that the relation is in fact exact. Interestingly, this relation signals an increase in transcendental weight (assigning weight 1 to ${\rm arcsinh}$) of the singular term between the 1SF and 2SF radial actions, hinting at a more complicated function space for the finite part of the 2SF radial action,
%
\begin{widetext}
\begin{align}
\tilde{I}_{r,5}^{2 {\rm SF,fin.}} = 16 c_\phi \Bigg[
& c^5_\phi  \left(
                {-}\frac{3}{160 \sigmabar^4}
                {+}\frac{1}{80 \sigmabar^3}
                {+}\cdots
            \right)
+ c^4_\phi \left(
               {-}\frac{4}{3 \sigmabar}
               {+}\frac{38}{45} 
               {+}\cdots
            \right)
+ c^3_\phi \left(
               \frac{1}{2 \sigmabar^2}
               {-}\frac{121}{18 \sigmabar}
               {+}\cdots
            \right)
 \\             
+&c^2_\phi \left(
               {-}\frac{12}{\sigmabar}
               {-}\frac{370}{9}
               {+}\cdots
            \right)
+c^1_\phi \left(
               {
               {-}102
               {+}\frac{232\sigmabar}{9}
               }
               {+}\cdots
            \right)
+c^0_\phi \left(
               {
                {-}72 \sigmabar
                {+}36 \sigmabar^2
                }
               {+}\cdots
            \right)            
\Bigg]. \nonumber
\end{align}
\end{widetext}
Interestingly, the relation~\eqref{eq:strange_relation} holds also for the coefficient of $c^5_\phi$ in $\tilde{I}_{r,5}^{2 {\rm SF,fin.}}$ and $\tilde{I}_{r,5}^{1 {\rm SF,fin.}}$, but not for the others.

To extract the scattering angle from the momentum-space radial actions $\tilde{I}^{{\rm nSF}}_{r,5}$, we Fourier transform the $\vect{q}$ dependence to impact parameter space (invert \eqref{eq:FT_J}), relate $b\equiv|\vect{b}|$ to the angular momentum via $b=J/|\vect{p}|$, and take a derivative with respect to $J$, see \eqref{eq:angle_Irad}. Following the same steps that led to Eq.~(5.44) in Ref.~\cite{Bern:2024adl}, we can write the SF-split 5PM scattering angle as
\begin{align}
\frac{\chi_5}{ m^2_1 m^2_2 } 
    =\frac{(m^4_1{+}m^4_2)}{m^2_1 m^2_2}\chi^{{\rm 0SF}}_{5}
        {+} \frac{(m^2_1{+}m^2_2)}{m_1 m_2}\chi^{{\rm 1SF}}_{5}
        {+}\chi^{{\rm 2SF}}_{5}\,.
\end{align}
In impact parameter space, the $\chi^{{\rm nSF}}_{5}$ are related to the momentum-space radial action coefficients via
\begin{align}
\label{eq:angle5PM}
 \frac{\chi^{{\rm nSF}}_{5}}{m^2_1 m^2_2} = \frac{G^5 \left(\mu^2 \tilde{b}^2\right)^{5 \epsilon}}{b^5} \frac{c^4_\phi}{|\vect{p}|^2 E}
 \Bigg[
 & -\frac{16}{\epsilon} \tilde{I}^{{\rm nSF,div.}}_{r,5} \\
 & \hspace{-2.5cm}+ \left(
     184 \tilde{I}^{{\rm nSF,div.}}_{r,5} 
    - 16 \tilde{I}^{{\rm nSF,fin.}}_{r,5}
   \right)
 \Bigg] + \mathcal{O}(\epsilon)\,, \nonumber
\end{align}
where $\tilde{b}^2 = b^2 \pi e^{\gamma_{\rm E}}$.  As in the radial action, the singular term should cancel once the 2SF radiation-reaction contributions to the scattering angle are included.

Our result passes several nontrivial checks. As stated above, at 0SF and 1SF, we agree with the expansion of our exact-in-velocity results of Ref.~\cite{Bern:2024adl}. 
Following the discussion in Ref.~\cite{Bern:2019crd}, the leading in velocity terms of the scattering angle are predicted by lower-order iterations. These velocity iterations arise from the integral representation of the scattering angle in terms of the radial momentum, see Eq.~(\ref{eq:angle_Irad}), and they determine the angle at orders $\sigmabar^{-j}$, $j=2,3,4,5$,
\begin{align}
    \hspace{-.3cm}
    \chi_5=\frac{16 \chi _1\chi_4 \left[1+\epsilon A\right]}{3 \pi }
    -\frac{9 \chi _1^5}{80} 
    -\frac{3}{2} \chi _3 \chi _1^2
    +\mathcal{O}\left(\frac{1}{\sigmabar},\epsilon\right),
    \hspace{-.3cm}
   \label{SmallVelPredict}
\end{align}
where we used the vanishing of the 2PM scattering angle, $\chi_2{=0}$, in $\mathcal{N}=8$ supergravity, and $A=\frac{55}{6}{-}16 \ln 2$. We checked (using the unpublished 4PM angle $\chi_4$~\cite{4PM-N8-unpublished}) that our results in Eq.~\eqref{eq:angle5PM} agree with this prediction. Note that the $\frac{1}{\epsilon}$ divergence of the 2SF angle is partially checked by the velocity iteration consistency at order $\sigmabar^{-2}$ and the presence of a similar divergence of the 4PM angle $\chi_4$.

\medskip
\Section{Discussion---}
%
%
\begin{figure}[t!]
\centering
\includegraphics[scale=.46]{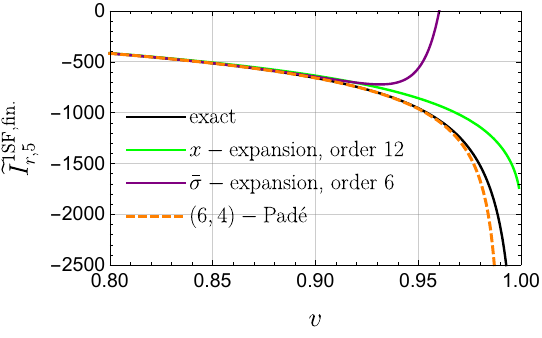}\  \ 
\includegraphics[scale=.46]{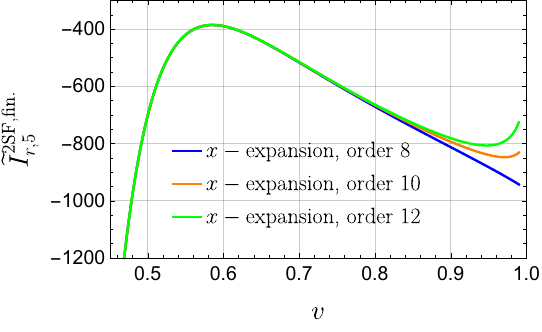}
\vspace{-.8cm}
\caption{\label{fig:RadialAction}
Potential region momentum-space radial actions in $\mathcal{N}=8$ supergravity at 5PM as a function of the velocity $v$ for $\cos\phi=0$. In the left panel, we show the exact finite part of the 1SF result of  Ref.~\cite{Bern:2024adl} compared to various series expansions. The right panel shows series expansions of the finite part of the 2SF radial action. 
}
\end{figure}
%
%
To gain insight into the quantitative features of our series solutions, in \fig{fig:RadialAction} we show the finite parts of the 1SF and 2SF momentum-space radial actions, which are directly related to scattering angles via Eq.~\eqref{eq:angle5PM}. We explore the robustness of different expansion schemes at 1SF order (left panel) and compare them to the exact velocity dependence available from Ref.~\cite{Bern:2024adl}. At 2SF order (right panel), the exact velocity dependence is not available, so we rely on different truncation orders to assess convergence. 

It is important to note that a judicious choice of variables makes the series solution converge over the entire physical region. So far, we have expressed results in terms of the relativistic factor $\sigma$ (or $\bar{\sigma}$), although the relative velocity $v$ (used in PN expansions) is another natural parameter near the static limit. Here, the radius of convergence of the series solution to the differential equations is set by the nearest singularity of the matrix $\hat{M}$, whose location depends on the chosen variables. Both the $\sigma$- and $v$-parameterizations introduce spurious poles that restrict convergence. However, in terms of the variable $x$, defined by $\sigma=(1+x^2)/(2x)$, these poles either lie outside the domain $|x-1|<1$ or fall within the physical region $x\in[0,1]$, where they are expected to cancel in observables. This indicates that, rather than expanding in terms of $\sigma$ or $v$, a more suitable choice is the variable $x$, expanded around the static limit $x=1$. Such an expansion is expected to converge across the full physical region. Current evidence indicates that this structure is more general, a conjecture that future studies of classical integrals at 6PM and beyond could substantiate.

Pad\'e approximants are well known to improve power-series expansions both in the construction of accurate waveforms~\cite{Damour:1997ub, Nagar:2016ayt, Nagar:2024oyk}, and in earlier applications in particle physics; e.g.~\cite{Bultheel:1984aa, Fleischer:1994ef, Fleischer:1996ju}. To gauge their effectiveness in our setting, we apply Pad\'e resummation to the Laurent-series expansion of the radial action, organizing terms by powers of $c_\phi$ and normalizing by $\sigmabar^4$ to obtain a power series through $\mathcal{O}(\sigmabar^{10})$. We then construct the Pad\'e approximant $R_{m,n}(\sigmabar) = P_m(\sigmabar)/Q_n(\sigmabar)$ with $(m,n)=(6,4)$, matched to this series. At 0SF order, the approximant reproduces the exact analytic result, while at 1SF order, it extends the agreement with the exact velocity result of  Ref.~\cite{Bern:2024adl} to higher $v$ (see left panel of Fig.~\ref{fig:RadialAction}), compared to a series expansion in $x$, and a series expansion in $\sigmabar$. In all cases, the agreement is better than 1\% for $v\lesssim 0.9$. The $\sigmabar$ expansion deviates significantly starting from $v=(2 \sqrt{2})/3\sim 0.94$ due to a singularity at $\sigma=-1$. At 2SF, performing a Pad\'e approximation leads to a spurious pole for certain choices of $(m,n)$ parameters, suggesting a more complicated structure. We therefore omit the 2SF Pad\'e approximation in the right panel of Fig.~\ref{fig:RadialAction} and show the convergence properties of the $x$ expansion for different truncation orders of the finite part of the 2SF radial action. This series exhibits rapid convergence for $v \lesssim 0.8$.  If complementary information on the asymptotic behavior of the radial action at $v{\to}1$ were available (following ideas on the high-energy resummation of Refs.~\cite{Rothstein:2024nlq, Barcaro:2025ifi}), one could anchor the approximants across the full kinematic range and further improve their accuracy, a direction we defer to the future.

%
\medskip
\Section{Conclusions---}
%
%
%
In this paper, we carried out the first 2SF computation in a gravitational theory, obtaining the potential-region scattering angle in $\mathcal{N}=8$ supergravity. Our calculation leveraged state-of-the-art Feynman integration tools, including new advances in IBP codes~\cite{FIRE7}, which should likewise benefit multi-loop collider physics.

After deriving the four-scalar amplitude integrand and IBP-reducing it, we solved the differential equations for all master integrals as a series expansion in $\sigmabar$ around the static limit $\sigmabar{\to} 0$ through $\mathcal{O}(\sigmabar^{15})$. These solutions determine the classical amplitude through 11 expansion orders; the loss of some orders is due to various poles in master coefficients. 
By comparing with 0SF and 1SF results where exact velocity is available, we find that significant improvements can be obtained via Pad\'e approximations. 
One may anticipate that combining the static-series solutions with complementary expansions around the high-energy limit ($\sigmabar \to \infty$) could yield reliable results throughout the entire kinematic range. Motivated by graviton dominance~\cite{tHooft:1987vrq, Muzinich:1987in, Parra-Martinez:2020dzs, Bern:2020gjj, DiVecchia:2020ymx}, the high-energy limit~\cite{Amati:1987wq, Amati:1987uf, Amati:1990xe} is also relevant for probing connections between different gravitational theories with classical scattering observables. We leave such explorations to future work. 

From a formal perspective, it would be interesting to promote our series solution to an exact analytic result, with complete control over (generalized) polylogarithms, elliptic integrals, and Calabi--Yau sectors. While we found evidence of nontrivial cancellations of elliptic and Calabi--Yau sectors in ${\cal N}=8$ supergravity, reminiscent of cancellations at 1SF order both in this theory~\cite{Bern:2024adl} and in Einstein's gravity~\cite{Driesse:2024xad}, a closed-form analytic 2SF calculation would confirm this. 

Although the present analysis was carried out in ${\cal N}=8$ supergravity, it is important to extend these results to Einstein's gravity. This entails more demanding IBP reductions than those performed here. Radiation modes would also need to be included. While complete physical results should be independent of a separation scheme between conservative and dissipative effects, a choice of scheme is necessary to compare with results of other computational frameworks~\cite{Wheeler:1949hn, Damour:1995kt, Damour:2016gwp, Barack:2018yvs, Driesse:2024xad}.
For improved waveform modeling, the PM results should then be incorporated into the EOB framework, as in, e.g., Refs.~\cite {Khalil:2022ylj, Buonanno:2024byg}, including a suitable analytic continuation scheme to bound orbits, for instance, along the lines of Refs.~\cite{Khalil:2022ylj, Bini:2024tft, Dlapa:2024cje, Dlapa:2025biy} or resummations incorporating the detailed singularity structure of observables~\cite{Damour:2022ybd, Long:2024ltn, OLtalk}. 

While the path to 2SF results at the fifth post-Minkowskian order in Einstein's gravity is nontrivial, it is increasingly clear that the remaining challenges can be overcome with existing ideas and methods. 

%
\medskip
\Section{Acknowledgements---}
%
%
We thank Alessandra Buonanno, Vittorio del Duca, Ira Rothstein, Michael Saavedra, Lorenzo Tancredi, and Stefan Weinzierl for helpful discussions.
We thank Julio Parra-Martinez and Chia-Hsien Shen for their permission to use unpublished 4PM results~\cite{4PM-N8-unpublished} in ${\cal N}=8$ supergravity. 
Z.B. and E.H.~are supported in part by the U.S. Department of Energy (DOE) under award number DE-SC0009937, and by the European Research Council (ERC) Horizon Synergy Grant “Making Sense of the Unexpected in the Gravitational-Wave Sky” grant agreement no. GWSky–101167314.
R.R.~is supported by the U.S.  Department of Energy (DOE) under award number~DE-SC00019066 and by a Senior Fellowship at the Institute for Theoretical Studies, ETH Z\"urich.
Part of the work reported in this paper was carried out while Z.B. and R.R. participated in the ``What is Particle Theory?'' program at the Kavli Institute for Theoretical Physics (KITP) and was supported in part by grant NSF PHY-2309135.
M.R. is supported by the Department of Energy (DOE), Contract DE-AC02-76SF00515. 
The work of  A.S. and V.S. was supported by the Moscow Center
for Fundamental and Applied Mathematics of Lomonosov Moscow State University under Agreement No. 075-15-2025-345.
M.Z.'s work is supported in part by the U.K.\ Royal Society through Grant URF\textbackslash R1\textbackslash 20109. For the purpose of open access, the author has applied a Creative Commons Attribution (CC BY) license to any Author Accepted Manuscript version arising from this submission.
The authors acknowledge the Texas Advanced Computing Center (TACC) at the University of Texas at Austin for providing high-performance computing resources that have contributed to the research results reported within this paper.  This research also used resources of the National Energy Research Scientific Computing Center (NERSC), a Department of Energy User Facility using NERSC award ERCAP 0034824. This work has also made use of the resources provided by the Edinburgh Compute and Data Facility (ECDF) (http://www.ecdf.ed.ac.uk/).
In addition, this work used computational and storage services associated with the Hoffman2 Cluster, which is operated by the UCLA Office of Advanced Research Computing’s Research Technology Group. We are also grateful to the Mani L. Bhaumik Institute for Theoretical Physics for support.


\bibliographystyle{apsrev4-1}
\bibliography{5PM_Neq8_2SF}

\end{document}